\documentstyle[12pt]{article}

\def\be{\begin{equation}}
\def\ee{\end{equation}}
\def\bea{\begin{eqnarray}}
\def\eea{\end{eqnarray}}
\def\a{\alpha}
\def\b{\beta}
\def\g{\gamma}
\def\l{\lambda}
\def\k{\kappa}
\def\z{\zeta}
\def\r{\rho}
\def\f{\phi}
\def\F{\Phi}
\def\kh{\frac{\k}{2}}
\def\lh{\frac{\l}{2}}
\newcommand{\ra}{\rightarrow}
\newcommand{\non}{\nonumber}
\newcommand{\pv}{\int_{-a}^{a}
\!\!\!\!\!\!\!\!\!-\!\!\!\!\!\!\!-\,\,\,\,}

\begin{document}

\begin{titlepage}
\begin{flushright}
KL --  TH -- 94/16   \\
hep-th/9407034
\end{flushright}

\vspace{1cm}

\begin{center}
{\LARGE ON THE SCHWINGER-DYSON EQUATIONS}     \\ \vspace{0.6cm}
{\LARGE FOR A VERTEX MODEL COUPLED TO} \\ \vspace{0.6cm}
{\LARGE 2D GRAVITY}  \end{center}

\vspace{.5cm}

\begin{center}
{\large Al.\ Kavalov}\footnote{E--mail: \ kavalov@physik.uni-kl.de}
\footnote{Permanent address: \ Yerevan Physics Institute,
Alikhanyan Brothers St. 2, Yerevan, 375036, Armenia} \\
{\it FB Physik, Universit\"at Kaiserslautern, 67663 Kaiserslautern,
Germany} \\   \end{center}
\vskip .2 cm

\vskip 1.5 cm

\begin{abstract}
\noindent
We consider a two matrix model with gaussian interaction
involving the term $tr ABAB$, which is quartic in angular variables.
It describes a vertex model (in particular
case - of F-model type) on the lattice of fluctuating geometry and is the
simplest representative of the class of matrix models describing  coupling
to two-dimensional gravity of general vertex models. This class includes
most of physically interesting matrix models, such as lattice gauge theories
and matrix models describing extrinsic curvature strings. We show that
the system of loop (Schwinger-Dyson) equations of the model decouples
in the planar limit and allows one to find closed equations for arbitrary
correlators, including the ones involving angular variables. This provides
a solution of the model in the planar limit.
We write down
the equations for the two-point function and the eigenvalue density and
sketch the calculation of perturbative corrections to the free case.
\end{abstract}


\end{titlepage}
\setcounter{page}{2}

\section{Introduction}

\setcounter{equation}{0}
\setcounter{footnote}{0}

The matrix model approach to 2D gravity (see \cite{GM, Ginsp}
for a review) has proven to be very useful and illuminating,
even providing a method of addressing the non-perturbative issues
of string theory \cite{Scaling}. It suffers, however, a severe
limitation, since all matrix models solved up to the present time
correspond to strings embedded in spacetimes of unphysical dimensions
$c\leq1$. The models with $c>1$ are investigated either perturbatively
\cite{Hik},
or numerically \cite{Num}.
The same $c=1$ - barrier appears in the continuum
approach to  non-critical strings too, making them inappropriate
for description of systems like gauge theories \cite{KPZ, Pol}.
Its physical origin is, at the present time, not very clear. There are
indications that it reflects an intrinsic instability of random surfaces
with respect to crumpling , is possibly connected
to the existense of the tachionic mode of the bosonic string ,
and can be understood in terms of condensation of spikes
\cite{Pol, DG, Spikes, Amb}.
These observations suggest that the corresponding phase transition may be
controllable in a model of surfaces with the action depending on the
extrinsic curvature \cite{ExCurv, Amb}. Being initially simpler, the
matrix model formulation of such models could shed a new light on the
problem and even suggest a way out of it, were it not for the technical
obstacle to solving interesting multi-matrix
models which  consists in inapplicability of the large-N saddle point technique
to the systems with the number of degrees of freedom of order $N^2$ ($N$
being the size of the matrix). The latter is not the case for one-matrix
models,
where the action does not depend on the angular part of the matrix and one
deals with  a system of $N$ eigenvalues. For the  next case, the
simplest two-matrix model,
Itzykson and Zuber were able to circumvent the problem by integrating
over the angles explicitly \cite{IZ}. The beautiful formula of
Itzykson and Zuber allows for the solution of multi-matrix
models with the tree-like structure of the target graph.
A direct further generalisation of this approach
to  target spaces containing cycles would invoke
calculation of the correlators of Itzykson-Zuber integral,  for which only
rather complicated  expressions are known \cite{MS} (see, however,
the interesting work of ref. \cite{Dalley2}).
     Given the importance of the matrix models, it is therefore of some
interest
to look for models which are tractable by methods not involving
the elimination of angular variables, such as the method of loop (Schwinger-
Dyson) equations (see \cite{M} for a review). The class of such models
includes, of course,
the ones solvable by previous techniques, in which case solving the loop
equations  sometimes allows the calculation of previously
unknown quantities, such as the correlators involving angular variables
\cite{AlfRet, Alf1, Staud}. The  interesting question is whether it is possible
to address
in this way any new models.
     In the present letter we describe a step in this direction. We define a
gaussian 2 matrix model describing coupling to gravity of a vertex model
(in particular case - of F-model type,  \cite{Baxter}).
The action of this model
contains a term $trABAB$ which prevents one from using the approach of
Itzykson and Zuber. However, the infinite system of planar loop
equations of the model can
be reduced to a closed set, thus giving an
analytical, though not very explicit, solution of the model and allowing one
to find, in principle, all planar correlators, including the ones involving
the angular variables. This provides a new example of a situation where
loop equations are more effective than the ``standard'' approach.

The model we consider belongs to a
very general class of matrix models describing vertex models
coupled to gravity. This class includes  most of the physically interesting
matrix models, among them lattice gauge theories and the matrix models
describing the extrinsic curvature strings. In view of the above,
the latter property seems
to be particularly interesting and is our primary motivation for
introducing and considering this class of models. We comment on this
connection in the next section, after having introduced our model which is,
unfortunately, only the simplest representative of the class and
corresponds to $c=1$. We hope, however, that it shares some
formal features with its more interesting relatives. It is interesting by
itself, too, as a novel $c=1$  two-matrix model.

     In the last section we use the orthogonal decomposition of the
matrices (used recently in the works \cite{AlfRet, Alf1, Alf2}) to derive
a system
of loop equations written in the variables, allowing for its
decoupling. We obtain a Fredholm equation of the second kind for
the eigenvalue density and a more general non-singular nonlinear
integral equation of Hammerstein's type for the general two-point
function $<trA^nB^m>$. This equations have unique solutions in terms
of Fredholm series. Analogous equations exist for arbitrary planar
correlators of the model. The equations we obtain are rather complicated,
but well suited for
perturbation expansion. We sketch the calculation in the first order.

\section{A model}

Consider a system of two hermitian matrices
with  the partition function defined by

\be
Z=\int dA dB e^{-NS},
\label{Z}
\ee
and the quadratic action
\be
S=tr \frac{r}{2} (A^2 + B^2) - 2c tr AB + \lh tr ABAB + \kh tr A^2B^2 .
\label{S}
\ee
This action leads to a perturbation theory involving two types of
propagators,
\bea
P(A,A) &=& P(B,B) = \frac{r}{r^2 - c^2},  \label{AA}  \\
P(A,B) &=&  \frac{c}{r^2 - c^2},                \label{BB}
\eea
and two types of vertices, with the couplings $\l$ and $\k$.
One can imagine labeling, for each vertex of the given diagram,
every A-type (B-type) leg by an outgoing (incoming) arrow. One then
obtains a configuration of a vertex model, which is very similar to the
familiar F version of the 6-vertex model \cite{Baxter}, which has a critical
regime with $c=1$.
The difference in the present case is that
the propagators of the type (\ref{AA}) allow
for configurations
with directions of the arrows not preserved along the link. This may be
viewed  as some deformation of the F-model. One can hope to recover the
pure F-model continuing the final expressions to the point $r=0$. Note,
however, that this limit is singular from the point of view of the original
integral (\ref{Z}). Alternatively, one can  put from the very beginning $c=0$.
This will leave the diagramms, containing only the  ``wrong" propagators
(\ref{AA}).
Than one can change locally the direction of all arrows for half of the
vertexes
to  show that there exists a configuration of the F-model corresponding to each
diagram. For some diagrams this procedure will meet, however, global
obstructions. Note also, that from the vertex model point of view the terms
$A^{4}$ and $B^{4}$ would correspond, if added to the action,  to deforming
the 6-vertex model to a non-critical 8-vertex model \cite{Baxter}. A way
to couple the F-model to gravity is to use the complex matrices
which give rise naturally to the oriented propagators (see \cite{Ginsp}).
This corresponds to taking $c = 0$ and adding the quartic terms
with the coefficients chosen to ensure the charge
conservation. Such a model arises also as a reduction of the
generalised Weingarten model and is solved in general, and, in the planar
case, in the presence of a vortex term, in the works \cite{Dalley1, Dalley2}.
\footnote{I thank S. Dalley for calling my attention to these works.}

The model (\ref{S}) belongs to a very general class of matrix models described
by
the actions
\be
S = \frac{1}{2} tr  l_{\a \b}  \F^{\a} \F^{\b}  +
V^{(3)}_{\a \b \g} \F^{\a} \F^{\b} \F^{\g}
  + \ldots
\ee
where $V_{\a \b \g \ldots}^{(n)}$  (with $V^{(2)} = l$)
are invariant with respect to cyclic transposition of indices.
In diagrammatic expansion of this model $l^{\a \b}$
( = inverse of $l_{\a \b}$) is the propagator of the matrix field
$\F$ and $V^{(n)}$ with $n > 2$ represent the vertexes.
{}From the point of view of the lattice dual to the given diagram
one has a statistical model with ``spins'' $\a$ living on the
links of a triangulation of some surface, and the statistical weight
obtained by prescribing the factor $V^{(n)}_{\a \b \g \ldots}$
to each face (triangle for $n=3$) and $l^{\a \b}$ to each
pair of the links identified in forming the surface.
This action was first written down by Bachas and Petropoulos in connection
to topological models on a two-dimansional lattice \cite{BP}.
The spins $\a$ take values in some set {\cal G}.
The familiar matrix models are the particular case with
$V_{\a \b \g \ldots}^{(0)} \sim \delta_{\a \b}
\delta_{\a \g} \ldots,  (n>2)$.
One can take
\be
[\Phi^{\alpha}]^{\dag} = \Phi^{\alpha ^{\dag}}   \non
\ee
with the dagger on the left hand side denoting the hermitian
conjugation and on the r.h.s. - some involution on {\cal G}. For
$\alpha ^{\dag} = \alpha$
one has a hermitian matrix model.

Let us show that the matrix models of this type describe, for
special choice of G, the extrinsic curvature strings.
To see the idea take $\alpha$ to be a pair (x,{\bf n}) where x is a
coordinate in 3-dimensional space and {\bf n} is a unit
3-vector, and arrange the couplings to have
\bea
S = \frac{1}{2} \sum_{x,{\bf n}_{1},{\bf n}_{2}} tr \F (x,{\bf n}_{1})
\F (x,{\bf n}_{2}) l_{{\bf n}_{1}{\bf n}_{2}} +
\sum_{x,{\bf n}}tr V(\F(x,{\bf n}))       \qquad \qquad                 \non
\\
+ \sum_{x_{1},x_{2},x_{3},{\bf n}} tr \F(x_{1},{\bf n})
\F(x_{2},{\bf n}) \F(x_{3},{\bf n}) \delta ({\bf n} \sim
{\bf x_{12}} \times {\bf x_{23}}) e^{-Area(123)}.
\eea
In the perturbation expansion of this model each link of the
triangulated surface dual to a diagram is labeled by a position
$x$ in the embedding space of the string and there is a unit vector
{\bf n} ascribed to each face of the triangulation. The $\delta$
-function in the cubic term ensures that {\bf n} is a unit normal
to the triangle (123), and one may take $l^{{\bf n}_{1},{\bf n}_{2}}
=exp(-{\bf n}_{12}^{2})$ to have this mutual orientation-dependent
factor associated with each neighbouring pair of triangles. The
$Area(123)$ denotes the area of the triangle (123), while the potential
term represents the couplings which add to the intrinsic area of
the surface leaving the extrinsic one unchanged,
and thus describe
the fluctuating intrinsic metrics (absent in the usual continuum
space formulation of the extrinsic curvature strings \cite{ExCurv}).

This action, though illustrating how one can hope to obtain
the matrix model describition of extrinsic curvature string,
is not the ``final'' correct version
of the extrinsic curvature matrix model. One still has to modify it by taking
the matrices to be complex and depending on some additional unit
vector. For the discrete embedding space this corresponds to
matrices living on the oriented links of the lattice and interacting
along the plaquettes in the usual Wilson way; the index {\bf n}
will describes in this case the orientation of the plaquette.
Identifying the matrices with different {\bf n} one gets the
Weingarten model \cite{Weing} (see \cite{Dalley2}).
Generalization to higher dimensions in obvious.
This connection of such matrix models to extrinsic curvature strings is
our primary motivation for their introduction and investigation.

Coming back to the model (\ref{S}), let us note that due to the
presence of the interaction term $tr ABAB$ the target graph of the model
contains cycles preventing one from using the approach of Itzykson and
Zuber. In spite of this, it is possible to proceed in calculating
arbitrary correlators of the model. We will show below that
the system of  the planar loop equations of the model reduces to a finite set.

\section{Loop equations}

It turns out to be convenient to write the loop equations in terms of the
variables defined via a decomposition
\be
M=\sum_{\alpha} m_{a \alpha} P_{a \alpha},        \label{Variables}
\ee
where $m_{a \alpha}$ are the eigenvalues and $P_{a \alpha}$
the projectors on the corresponding eigenvector. The latter satisfy
the conditions
\be
P_{a \alpha}P_{a \beta} = \delta_{\alpha \beta}P_{a \alpha},\quad
trP_{a \alpha}=1,\quad
\sum_{\alpha}P_{a \alpha}=1.  \quad                     \label{Proj}
\ee
We will denote below the eigenvalues of $A$ (resp. $B$) by $x_{\a}$
(resp. $y_{\b}$) and the corresponding projectors as $A_{\a}$
and $B_{\b}$. Another useful notation is
\be
P_{\a_{1} \b_{1} \a_{2} \b_{2} \cdots \a_{n} \b_{n}} =
tr A_{\a_{1}}B_{\b_{1}}
A_{\a_{2}}B_{\b_{2}} \cdots A_{\a_{n}}B_{\b_{n}},        \non
\ee
and also
\be
P_{n m \alpha_{2} \beta_{2} \cdots \alpha_{k} \beta_{k}}  =
\sum_{\alpha_{1} \beta_{1}} x_{\alpha_{1}}^{n} y_{\beta_{1}}^{m}
A_{\alpha_{1} \beta_{1} \alpha_{2} \beta_{2} \cdots \alpha_{k} \beta_{k}}.
\label{P}
\ee
To avoid the possible confusion let us
keep in mind that the numbers in the sequence of indices of the objects like
$P_{1 2\alpha_{2} \beta_{2} \cdots \alpha_{n} \beta{n}}$ will always
denote the power of the matrix in the corresponding position in the trace
in (\ref{P}),
while the Greek letters will be used for numbering the projectors. The same
will apply to the continuum notation $P(1,2,x_{3},y_{4}, \cdots, x_{n}, y_{n})$
with continious variables $x_{i}$ ($y_{i}$)  taking the places of the
discrete indexes $\alpha_{i}$ ($\beta_{i}$). The variables (\ref{Variables})
were recently used by Alfaro to solve the 2-matrix model and to write
down the loop equations for D-dimensional matrix models \cite{Alf1, Alf2}.

The equations we need follow from identities of type
\be
\int \frac{\partial}{\partial A_j^i}(A_{\a_{1}}B_{\b_{1}}
A_{\a_{2}}B_{\b_{2}} \cdots A_{\a_{n}}B_{\b_{n}})^{i}_{j}
 e^{-NS} = 0.
 \label{Generator}
\ee
In particular, one finds for $n=2$
\bea
& &  [ rx_{\a_1} - c y_{\b_2} + \kh x_{\a_1} y_{\b_2}^{2} ]
P_{\a_1 \b_1 \a_2 \b_2}
-\frac{1}{N} \sum_{\g \neq \a_1} \frac{1}{x_{\g \a_1}}(
P_{\a_1 \b_1 \a_2 \b_2}  +  P_{\g_{1} \b_{1} \a_{2} \b{2}} )       \non  \\
& & - \frac{1}{N}\sum_{\g \neq \a_{1}}
\frac{1}{x_{\g \a_1}} P_{\a_{1} \b_{1}} P_{\g \b_{2}}
(\delta_{\a_{1} \a_{2}} -  \delta_{ \g \a_{2}})
+ \l y_{\b_2} P_{\a_{1} \b_{1} \a_{2} \b_{2} 1 1}  +
\kh P_{\a_{1} \b_{1} \a_{2} \b_{2} 1 2} = 0.
\label{Pabab}
\eea
Summing this expression over $\alpha_{2}$ and using (\ref{Proj}) one obtains
the
relation for $n=1$:
\be
[ rx_{\a} - c y_{\b} + \kh x_{\a} y_{\b}^{2} ] P_{\a \b}
-\frac{1}{N}\sum_{\g \neq \a} \frac{1}{x_{\g \a}}(
P_{\a \b} +P_{\g \b})  +
\l y_{\b} P_{\a \b 1 1}  +  \kh P_{\a \b 1 2} = 0.
\label{Pab}
\ee
An analogous set of equations is obtained by taking in (\ref{Generator})
the derivative
$\frac{\partial}{\partial B}$ instead of $\frac{\partial}{\partial A}$.
This new equations can be taken into account by requiring
$P_{\alpha_1 \beta_1 \alpha_2 \beta_2 \cdots \alpha_n \beta_n}$
to be invariant under the cyclic transposition of indexes.

In the limit of $N \ra \infty$ we introduce the usual rescaled
variables $x_{i} = x_{\alpha}$, $ i = \frac{\alpha}{N}$,
$y_{j} = y_{\beta},  j = \frac{\beta}{N}$
and assume that the eigenvalue density $\r (x) = \frac{di}{dx}$
(the same for both matrices) has a finite
support $[-a, a]$. One obtains, for instance, for eq.(\ref{Pab}):
\bea
[rx - c y + \kh x y^{2} + \pv \frac{u(z)dz}{z-x}] P(x,y)
+ \pv\frac{u(z)dz}{z-x}P(z,y)  \non                \\
+ \l y P(x,y,1,1) + \kh P(x,y,1,2) = 0,
\label{Pxy}
\eea
with the density of eigenvalues and the correlators normalized
by the conditions following from (\ref{Proj})
\bea
\int_{-a}^{a} \r(x) dx &=& 1,   \\
\int_{-a}^{a} \r(x) P(x,y) dx &=& 1, \\
\int_{-a}^{a} \r(x) P(x,y,1,1) dx &=& y  P(1, y),  \quad  etc.
\eea
The symbol $\displaystyle{\int
\!\!\!\!\!\!-\!\!\!\!\!-}$ denotes the principal value of the integral.
The equation (\ref{Pabab}) is rewritten analogously as
\bea
&  & [rx_1- c y_2 + \kh x_1 y_2^{2} + \pv\frac{u(z)dz}{z-x_1}]
         P(x_1,y_1,x_2,y_2) + \pv\frac{u(z)dz}{z-x_1}P(z,y_1,x_2,y_2)
\non   \\
&  &   + P(x_1,y_1)  \pv\frac{u(z)dz}{z-x_1}P(z,y_2)
         [\frac{\delta (x_1 - x_2)}{\r (x_1)} -  \frac{\delta (z - x_2)}{\r
(z)}] \non  \\
& &   + \l y P(x_1,y_1,x_2,y_2,1,1) + \kh P(x_1,y_1,x_2,y_2,1,2) = 0.
\label{Pxyxy}
\eea

Taken for arbitrary $n$, (\ref{Generator}) thus provides an infinite set of
linear
relations between the correlators of the system.
Our aim is to show, that eqs.(\ref{Pxy}) and (\ref{Pxyxy}) reduce, in fact, to
a closed
system.
To see this, note that multiplying both sides of eq.(\ref{Pxy}) by $\r(x)$
and integrating over $x$ one obtains an algebraic equation for the
correlator $P(x,1)$ which gives
\be
P(x,1) = \frac{cx}{r + (\lambda + \kappa)x^{2}}.         \label{P1x}
\ee
A similar thing occures when one integrates over $x_{1}$ in eq.(\ref{Pxyxy}).
One
obtains the equation
\bea
[r + \lambda y_{1} y{2}  + \frac{\kappa}{2} (y_{1}^{2} + y_{2}^{2})]
        P(1, y_1,x_2,y_2) =  \non \qquad \qquad \qquad \qquad   \qquad      \\
      c \frac{\delta (y_{1} - y{2})}{\rho(y_{1})} P(x_{2},y_{2})
- \pv\frac{dz \r(z)}{z-x} [P(x,y_1) P(z,y_2) + P(z,y_1)P(x,y_2)],     \qquad
\label{P1yxy}
\eea
which expresses the  4-point correlator $P(1,y_{1}, x_{2},y_{2})$ in terms of
2-point functions.  From here one finds easily
\bea
  P(x,y,1,1) = \frac{cy^2}{r + (\l + \k) y^2} P(x,y) \non \qquad \qquad \qquad
\qquad    \qquad   \qquad   \qquad    \\
  - \int_{-a}^a \frac{d \z \r (\z) \z}{r + \l \z y + \kh (\z^2 + y^2)}
         \pv\frac{\r(z)dz}{z-x} [P(x,\z)P(z,y) + P(x,y)P(z,\z)],
\qquad  \qquad
\label{P11xy}
\eea
and
\bea
 P(x,y,1,2) = \frac{cy^3}{r + (\l + \k) y^2} P(x,y) -  \non    \qquad   \qquad
  \qquad   \qquad  \qquad \\
   \int_{-a}^a \frac{d \z \r (\z) \z^2}{r + \l \z y + \kh (\z^2 + y^2)}
         \pv\frac{\r(z)dz}{z-x} [P(x,\z)P(z,y) + P(x,y)P(z,\z)].
\label{P12xy}
\eea
Substituting these into eq.(\ref{Pxy}) gives a non-linear singular integral
equation involving two unknown functions, $\r(x)$ and $P(x,y)$.
The second equation is provided by the symmetricity condition $P(x,y) =
P(y,x)$.

     Let us now rewrite this system  resolving the singular
part of the equations. We introduce the functions
\bea
H(x)   &=& \int_{-a}^{a} \frac{u(z)dz}{z-x},                       \\
\f(x;y) &=&   \int_{-a}^{a}\frac{u(z)dz}{z-x}P(z,y),
\eea
and
\be
G(x;y)   =   rx - cy - \frac{\kappa}{2} x y^2 + H(x).
\ee
The variable $x$ in these functions takes values on the complex plane
cut along the interval $[-a,a]$, while $y$ is real and belongs to $[-a,a]$.
One has, for $x \in [-a,a]$,
\be
Disc_x\f(x;y) \equiv \f(x+i0;y) - \f(x-i0;y) = 2\pi i \r(x) P(x,y)
\ee
and
\be
Cont_x\f(x;y) \equiv \f(x+i0;y) + \f(x-i0;y) =  2 \pv\frac{\r(z)dz}{z-x}P(z,y).
\ee
Using these relations, in (\ref{Pxy}), (\ref{P11xy}) and (\ref{P12xy})
one  can easily show  that the both
sides of the equation (\ref{Pxy}) are given by discontinuities of some
functions.
One obtains
\bea
& & Disc_x  G(x;y)\f(x;y) = - Disc_x  (\l + \kh) \frac{cy^3}{r + (\l + \k) y^2}
\f(x,y)
\non  \\
& & + Disc_x  \f(x,y) \int_{-a}^a d\z \r(\z) \frac{\l \z y + \kh \z^2}
{r + \l \z y + \kh (\z^2 + y^2)} \f(x,\z),
\eea
which implies that
\bea
& & G(x;y)\f(x;y) + (\l + \kh) \frac{cy^3}{r + (\l + \k) y^2} \f(x,y)   \non
\\
& &  - \f(x,y) \int_{-a}^a d\z \r(\z) \frac{\l \z y + \kh \z^2}
{r + \l \z y + \kh (\z^2 + y^2)} \f(x,\z) = Pol.(x),
\label{Pol}
\eea
where $Pol.(x)$ is some polynomial of x. Taking
$x \ra \infty$ and using the known behaviour
of $H(x)$ and $\f(x,y)$ in this limit, one finds $Pol.(x)=
r - \frac{\k}{2}y^2$ and, after some algebra, (\ref{Pol}) reaches its final
form
\be
\f(x;y) K \f(x;y) + [x - \frac{cy}{r + (\l + \k) y^2}] \f(x;y) + 1 = 0
\label{Fixy}
\ee
with
\be
K \f(x;y) = \int_{-a}^a d\z \r(\z) \frac{\f(x,\z)}{r + \l \z y + \kh (\z^2 +
y^2)}.
\ee
This is a non-linear integral equation of Hammerstein's type
which defines the function $\r(x) \f(x,y)$. Another equation,
expressing the symmetricity of $P(x,y)$ can be written as
$\F(x,y) = \F(y,x)$ with
\be
\F(x,y) =  \int_{-a}^{a}\frac{\r(\z)d\z}{\z-y} \f(x,\z).
\label{Sym}
\ee
Taking here the limit of $y \ra \infty$ one obtains the simpler
condition
\bea
H(x) &=& \int_{-a}^a d\z \r(\z) \f(x,z)  \non  \\
        &=& - \int_{-a}^a \frac{dz \r(z)}{x - \frac{cz}{r + (\l  + \k) z^2} +
        \int_{-a}^a \frac{d\z \r(\z) \f(x,\z)}{r + \l \z y + \kh (\z^2 + y^2)}}
{}.
\label{sym}
\eea

     Equations (\ref{Fixy}) and (\ref{Sym}) allow one to find
the eigenvalue density $\r(x)$ and the two-point function $P(x,y)$,
and provide the decoupling of the system (\ref{Generator}). The existence of
solutions for (\ref{Fixy}) and (\ref{Sym}) follows from the theorems of the
theory of integral equations, but we do not know, at the present
moment, how to find them explicitly. The same theorems say a lot about
the properties of the solutions. Leaving the complete investigation of
the system (\ref{Fixy}, \ref{Sym}) to the future, let us have a brief look on
its
perturbative properties.

Consider first the free case.
Taking $\l = \k =0$ one finds from the symmetricity condition (\ref{sym})
\be
\frac{1}{r}H_0(x) = - \int_{-a}^{a} \frac{\r(z)dz}{rx - cz + H_0(x)},
\ee
or
\be
H_0 \left(\frac{rx + H_0(x)}{c}\right) = \frac{c}{r}H_0(x).
\ee
This functional equation is similar to the one found in different contexts by
Matitsyn \cite{Mat} and by Boulatov \cite{B}. It can be solved either by
noticing directly that it implies
\be
r H_0(x)^2 + (r^2 - c^2) x H_0(x) = const.
\ee
and finding, for $x \ra \infty$, the constant to be
$c^2 - r^2$, or by taking $c \ll 1$ and performing
successive approximations. The solution is given by
\be
H_0(x) = -\frac{r^2 - c^2}{2 r} \left[ x - \sqrt{x^2 - \frac{4r}{r^2 - c^2}}
\right]
\ee
and leads, of course, to Wigner's semi-circle distribution
\be
\r_0(x) =  \frac{r^2 - c^2}{2 \pi r} \sqrt{x^2 - \frac{4r}{r^2 - c^2}} .
\ee
The 2-point function $\f$ is, in this limit, given by
\be
\f_0(x;y) = - \frac{r}{rx - cy - H_0(x)} .
\ee

Coming back to eq. (\ref{Fixy}) note that in the limit of
$x \ra \infty$ the non-linear term in it is much
smaller that the other two. Applying the successive approximations to
\be
\f(x;y) = -  \frac{1}{x - \frac{cy}{r + (\l + \k) y^2}}
-   \frac{1}{x - \frac{cy}{r + (\l + \k) y^2}} \f(x;y) K\f(x;y),
\ee
one obtains $\f(x,y)$ as an expansion in $\frac{1}{x - \frac{cy}
{r + (\l + \k) y^2}}$. The first term is given by
\be
\f^{(0)}(x;y) = -  \frac{1}{x - \frac{cy}{r + (\l + \k) y^2}},
\ee
while the succesive ones contain $\r(x)$:
\be
\f^{(1)}(x;y) = -  \frac{1}{x - \frac{cy}{r + (\l + \k) y^2}}
-  \frac{1}{x - \frac{cy}{r + (\l + \k) y^2}} \f^{(0)}(x;y)
K\f^{(0)}(x;y),\quad    etc.
\label{Iterations}
\ee
Comparing (\ref{Iterations}) to the expansion
\be
\f(x;y) = - \frac{1}{x} - \frac{P(1,y)}{x^2} - \frac{P(2,y)}{x^3} - \cdots,
\ee
one finds the correlators $P(n,y)$ in terms of $\r $.
In particular,
\be
P(2,y) =   - \frac{c^2 x}{r + (\l + \k)x^2}
+ \int_{-a}^a \frac{d\z \r(\z)}{r + \l \z y + \kh (\z^2 + y^2)}.
\label{P2x}
\ee
Note an exact relation following from (\ref{Iterations}):
\be
\f(x;y;c) = \f(x - \frac{cy}{r + (\l + \k) y^2}; y; c=0).
\ee

Finally, multiply both sides of eq.(\ref{Pxy}) by $\r(y)$
and integrate over $y \in [-a,a]$. One obtains
\be
\pv \frac{\r(z)}{z-x} = -\frac{1}{2} \left[ rx -cP(x,1) + \k x P(x,2)
+ \l P(x,1,1,1) \right].
\ee
Using (\ref{P11xy}) to find
\be
P(x,1,1,1) =  \frac{c^2 x^3}{[r + (\l + \k)x^2]^2}
+  \int_{-a}^a \frac{d\z \r(\z) \z}{r + \l \z y + \kh (\z^2 + y^2)},
\ee
as well as (\ref{P2x}) and (\ref{P1x}), one arrives at the equation
containing $\r(x)$ as the only unknown:
\be
\pv\frac{\r(z)}{z-x}= \frac{1}{2} f(x),
\label{rho}
\ee
where
\be
f(x) = - rx-\frac{r c^2x}{[ r + (\l + \k)x^2]^2}+ \int_{-a}^a d\z \r(\z)
\frac{\l \z + \k x}{r + \l \z x + \kh (\z^2 + x^2)} .
\ee
The same equation could, of course, be derived directly by
integrating in (\ref{Z}) over one of the matrices.
This would lead to an effective
action involving `non-local' interactions of the type $(tr A^n)^m$.
Such matrix models were considered  in the works of ref.\cite{Wadia}.

Once again, one can resolve the singular part of
(\ref{rho}) to reduce it to
\bea
H(x) = r \frac{\sqrt{a^2-x^2}}{2\pi i} \int_{-a}^{a}\frac{dz}{z-x}
\frac{1}{\sqrt{a^2-z^2}} \left[-z+\frac{c^2z}{[r + (\l + \k)z^2]^2} \right]
\non  \\
- \frac{\sqrt{a^2-x^2}}{2\pi i} \int_{-a}^{a}\frac{dz}{z-x}
\frac{1}{\sqrt{a^2-z^2}}  \int_{-a}^{a} d\z \r(\z)
\frac{\l \z + \k z}{r + \l \z z + \kh (\z^2 + z^2)} .
\label{H}
\eea
The first integral in this equation can be calculated easily by rewriting it
as a contour integral around the cut and inflating the contour to
touch the singularities at $z=x,$  $\pm i \sqrt{r/(\l + \k)}$ and $\infty$.

Taking the discontinuity of both sides of (\ref{H}) one
obtains a Fredholm equation of second kind for $\r(x)$.  One can
then perform the succesive approximations  once again to obtain
the solution in terms of Fredholm series.
Alternatively, one can  use (\ref{H}) directly to calculate $H(x)$ as
power series in $\l$ and $\k$. Consider, as a simple example,
the case of $c=\k=0$.
One finds from (\ref{H}), in the lowest order
\be
H(x) = -\frac{1}{2} \left[r - (\frac{\l}{\r})^2 \r_2 \right] \left[x -
\sqrt{x^2 - a^2}
\right],
\label{Pert}
\ee
where the coefficient $\r_2 = \int_{-a}^{a}dz \r(z) z^2$, as well as $a^2$,
have to be determined from comparison of (\ref{Pert}) to the expansion
\be
\r(x) = - \frac{1}{x}  - \frac{\r_2}{x^3}  - \frac{\r_4}{x^5} - \cdots
\ee
One finds, after some algebra,
\bea
\r_2 &=& \frac{a^2}{4},                    \non  \\
a^2 &=& \frac{4}{r} \left[1 +  (\frac{\l}{\r})^2 \frac{1}{r^2} \right].
\eea
It is not difficult to write down the corrections of first two orders
for general $c$. The formulae are, however, not very illuminating.

\section{Comments}

This concludes our preliminary investigation of the model (\ref{S}).
Evidently, there is still much to be done, as the progress achieved in
the present work is mostly of technical nature. We feel however that
it is of some interest, as a demonstration of usefulness of the
loop equations for calculations of the correlators involving
angular variables for the vertex model on the lattice of fluctuating
geometry which was not considered previously.
     Our interest in the model (\ref{S}) was triggered by the observation
that it contains an interaction of the same type as the matrix models
describing extrinsic curvature strings. The latter models can lead
to important physical insights. Further work is in progress.

\subsection{Acknowledgements}

The author is grateful to J. Amb{\o}rn, S.~Cordes, A. Ganchev,
O. Lechtenfeld, V.F. M\"uller, W. R\"uhl and especially to D. Boulatov,
M. Douglas and B. Rusakov for useful and stimulating discussions.


\end{document}